\documentclass[aps,prl,showpacs,twocolumn,amsmath]{revtex4}
\usepackage{graphicx}

\newcommand{\bq}{\begin{equation}}
\newcommand{\eq}{\end{equation}}
\newcommand{\bn}{\begin{eqnarray}}
\newcommand{\en}{\end{eqnarray}}
\newcommand{\bsub}{\begin{subequations}}
\newcommand{\esub}{\end{subequations}}

\newcommand{\eps}{\epsilon}

\begin{document}
\title{Breaking of Phase Symmetry in Non-Equilibrium Aharonov-Bohm Oscillations through a Quantum Dot}
\author{Vadim Puller$^1$}
\author{Yigal Meir$^{1,2}$}
\author{Martin Sigrist$^3$}
\author{Klaus Ensslin$^3$}
\author{Thomas Ihn$^3$}
\affiliation{$^1$Department of Physics, Ben-Gurion University of
the Negev, Beer Sheva 84105 Israel\\$^{2}$ The Ilse Katz Center
for Meso- and Nano-scale Science and Technology, Ben-Gurion
University, Beer Sheva 84105, Israel\\$^{3}$ Solid State Physics
Laboratory, ETH Z\"{u}rich, 8093 Z\"{u}rich, Switzerland}

\begin{abstract}
{Linear response conductance of a two terminal Aharonov-Bohm (AB)
interferometer is an even function of magnetic field.
This {\em phase symmetry} is no expected to hold beyond the linear response regime.
In simple AB rings the phase of the oscillations changes smoothly (almost linearly) with voltage bias. However, in an interferometer with a quantum dot in its arm, tuned to the Coulomb blockade regime, experiments indicate that phase symmetry seems to persist even in the nonlinear regime.

In this letter we discuss the processes that break AB phase symmetry. In particular we show that breaking of phase symmetry in such an interferometer is possible only after the onset of inelastic cotunneling, i.e. when the voltage bias is larger than the excitation energy in the dot. The asymmetric component of AB oscillations is significant only when the contributions of different levels to the symmetric component nearly cancel out (e.g., due to different parity of these levels), which explains the sharp changes of the AB phase. We show that our theoretical results are consistent with experimental findings. }
\end{abstract}
\date{\today}

 \maketitle

The Aharonov-Bohm (AB) effect allows for studying the transmission phase
through a mesoscopic structure, e.g. a quantum dot (QD),  by placing it
in one of the arms of an AB interferometer \cite{Yacoby,Schuster}.
In a two terminal interferometer the phase of the AB oscillations in the
linear response
conductance can only assume the values $0$ or $\pi$ (i.e. the oscillations
have either maximum or minimum at zero magnetic field), even though the
transmission phase through the QD
can change continuously. This \emph{phase symmetry}, i.e. the property
that the linear response conductance of a two-terminal device is an even
function of magnetic flux,
can be understood within a one-particle
picture \cite{LeviYeyati} and is, in fact, a
manifestation of more general linear-response Onsager-B\"{u}ttiker
symmetries \cite{Onsager,Buttiker}.
Deviations from phase symmetry in two-terminal devices
 in the nonlinear regime have been studied theoretically
\cite{Bruder,Buttiker2,Spivak}, as well as in experiments on AB cavities
\cite{PRexperiments} and AB rings \cite{electricAB}. The resulting phase of
the AB oscillations changes smoothly (almost linearly) with increasing voltage
bias \cite{electricAB}.

Rather puzzlingly, a recent experiment \cite{Ensslin},
which studied a voltage-biased AB interferometer with Coulomb blockaded
QDs in its arms, observed AB oscillations which remained practically symmetric.
The phase of the oscillations changed with voltage bias $V$ in a highly
non-monotonous fashion: it remained close $0$ and $\pi$, but switched abruptly
between these two values as a function of the bias
voltage, with the first switching occurring
when the voltage about equal to the level spacing to the first excited state
$\Delta$, i.e. near the onset of inelastic cotunneling.

Indeed, breaking of the phase symmetry in the regime of inelastic cotunneling
have not been addressed theoretically thus far. In particular, presence of
the finite bias threshold for the inelastic cotunneling renders inapplicable
the methods
based on expansion in powers of of the bias voltage \cite{Buttiker2}, and
thus cannot explain the experimental observations.
In this Letter we address the phase asymmetry of AB oscillations
in a QD interferometer with a Coulomb blockaded dot by systematically
analyzing transport processes of different order in lead-to-lead tunnel
coupling. Based on their dependence on voltage
bias and magnetic field we establish that the bias
dependence of the AB phase is highly non-monotonous. In particular,
(i) the oscillations indeed
remain symmetric up to the onset of inelastic cotunneling ($eV\simeq\Delta$)
(i.e. with AB phase $0$ or $\pi$), in agreement with experiments;
(ii) with onset of inelastic cotunneling, AB oscillations acquire
non-zero asymmetric component, which however is usually smaller than the
symmetric component, the oscillations thus remaining
nearly symmetric; (iii) the asymmetric
component may become dominant, if the contributions of different levels
to even AB oscillations nearly cancel out (e.g., due to different parity of
these levels) \cite{PullerMeir}.
The theoretical findings are supported by the in-depth analysis
of the experimental data of Ref. \onlinecite{Ensslin}.

\begin{figure}[tbp]
  \includegraphics[width=3.3in]{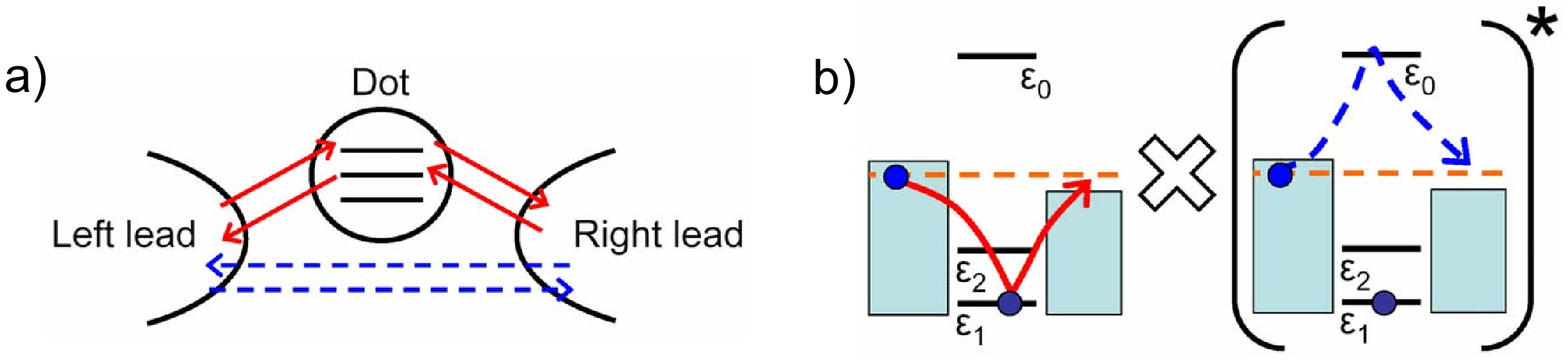}
  \caption{(color online) (a) Schematic representation of the device studied
  in this paper. Solid red and dash blue arrows show
  cotunneling processes and direct lead-to-lead tunneling
  respectively. (b) Example of a lowest order cotunneling process
  contributing to AB oscillations.}\label{fig:device}

\end{figure}

\paragraph{Theoretical formulation}
We consider an AB interferometer schematically shown in
Fig.~\ref{fig:device}a. One arm of the interferometer contains a QD which is
assumed to be in Coulomb blockade regime. The current can flow either by
means of cotunneling via the QD or by direct lead-to-lead tunneling through
the open arm of the interferometer \cite{onedot}, whereas the number of
electrons occupying the QD does not change.

We describe the system by Hamiltonian
$H=H_L+H_R+H_D+V+W$,
where $H_\mu=\sum_{E}Ec_{\mu E}^+c_{\mu E}$
is the Hamiltonian of electrons in lead $\mu=L,R$; $E$ labels energy states
within one lead.  $H_D=\sum_{\beta}\eps_{\beta}d_{\beta}^+d_{\beta}$ is
the Hamiltonian of
the QD, which contains only one electron and has energy levels
$\eps_{\beta}$. $c_{\mu E}$ destroys a lead electron in state $\mu E$,
$d_{\beta}$ destroys QD state $\beta$ \cite{Supplement}.

$W$ and $V$ describe, respectively, electron
transitions between the leads through the open arm or through the arm that
contains the QD. Due to the Coulomb blockade, the number of electrons in
the QD after the electron transfer remains unchanged, but the process can
be accompanied by change of the QD state. These terms in the Hamiltonian
are given by
\bsub\begin{eqnarray}
 W&=&\sum_{\mu E}\sum_{\mu'\ E'}
W_{\mu;\mu'}e^{i\phi_{\mu\mu'}}\
c_{\mu E}^+ c_{\mu'E'}\\
V&=&\sum_{\beta,\beta'}\sum_{\mu E}\sum_{\mu' E'}
V_{\mu;\mu'}^{\beta;\beta'}\
d_{\beta}^+c_{\mu E}^+ c_{\mu'E'} d_{\beta'},
\end{eqnarray}\esub
where $W_{\mu;\mu'}$, and $V_{\mu;\mu'}^{\beta;\beta'}$ are real, and $\phi$ is the magnetic flux through the interferometer ($\phi_{RL}=-\phi_{LR}=\phi$, $\phi_{LL}=\phi_{RR}=0$) \cite{Supplement}.

\begin{figure}[tbp]
  \includegraphics[width=3.5in]{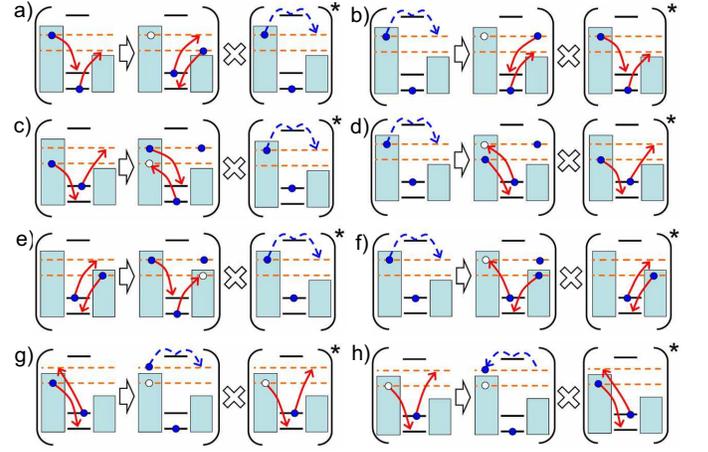}
  \caption{(color online) Examples of different pairs of third order
  processes: (a) and (b) (or (c) and (d)) are two processes whose
  contributions to odd AB oscillations  mutually cancel out; process
  (a) is elastic, whereas (b) is inelastic; the QD is initially in
  its {\em ground} ({\em excited}) state. (e) (or (g)) is an example
  of an {\em elastic} ({\em inelastic}) third order process which
  gives non-zero contribution to the odd AB oscillations. The other
  process constructed from the same matrix elements and beginning from
  the same initial state, (f) (or (h)), does not contribute to the
  current.}\label{fig:third_order}
\end{figure}

\paragraph{Breaking of phase symmetry}
It is easy to see that the second order processes contributing to the
AB oscillations (which necessarily involve one tunneling amplitude through
the open arm, $W$, and one through the dot, $V$), such as the one depicted
in  Fig.~\ref{fig:device}b (where $\eps_0$ represents the open arm), are
necessarily symmetric with respect to magnetic field. The asymmetric AB
oscillations appear when we account for higher order tunneling processes.
Typical third-order contributions to AB oscillations are depicted in
Fig.~\ref{fig:third_order}. As an example, the probabilities of the
processes shown in Fig.~\ref{fig:third_order}a,b, are, respectively,
\bsub
\bn 4\pi\Re\left[\frac{\left(W_{R;L}e^{i\phi}\right)^*
V_{R;R}^{1;2}V_{R;L}^{2;1}}
{\eps_{1}+E_L-\eps_{2}-\tilde{E}_R+i0^+}\right]\delta(E_L-E_R),\label{eq:elastic}\\
4\pi\Re\left[\frac{\left(V_{R;L}^{2;1}\right)^*
V_{R;R}^{2;1}W_{R;L}e^{i\phi}}
{E_L-E_R+i0^+}\right]\delta(E_L+\eps_{1}-\tilde{E}_R-\eps_{2}), \label{eq:inelastic}
\en\label{eq:proc}
\esub
($\Re$ represents the real part). These factors consist of the second order
tunneling amplitude (which contains the energy denominator) multiplied by
the complex conjugate of the first order tunneling amplitude: this is
reflected in the obvious fashion in Fig.~\ref{fig:third_order}, upon which
the following discussion is built. There are also processes (not shown here)
in which instead of an electron one considers tunneling of a hole.

In order to obtain correction to the current, the probabilities in
Eq.~(\ref{eq:proc})  are multiplied by the factor
$P_{1}f_{L}(E_L)[1-f_{R}(E_R)] [1-f_{R}(\tilde{E}_R)]$ (which also limits
possible intermediate  states) and integrated over $E_L,E_R$ and $\tilde{E}_R$.

The contribution to AB oscillations coming from the real parts of the
denominators in Eqs.~(\ref{eq:proc}) is even in magnetic field.
Asymmetric terms may result from the imaginary part of the denominators
in Eqs. (\ref{eq:proc}), which we treat according to prescription
$1/(E+i0^+)=1/E-i\pi\delta(E)$~\cite{HolsteinOra}. The delta-function means
that
the contribution to AB oscillations odd in magnetic field may result
only from the processes in which the {\em intermediate
state lies on
the same energy shell with the initial and the final states}, which
for our example means that $E_L+\eps_{1}=E_R+\eps_{1}=\tilde{E}_R+\eps_{2}$.
This is the case shown in all our figures.

The asymmetric contribution due to the process (\ref{eq:proc})a is thus
given by
\begin{equation}
-2W_{R;L}V_{R;R}^{1;2}V_{R;L}^{2;1}
\delta(\eps_{1}+E_L-\eps_{2}-\tilde{E}_R)\delta(E_L-E_R)\sin\phi.\label{process}
\end{equation}
On the other hand, the asymmetric contribution of the process (\ref{eq:proc})b
is given by exactly the same expression, but with the opposite sign, and thus
the asymmetry contribution is canceled between these two processes. This is
no surprise. The first process (Fig.~\ref{fig:third_order}a) corresponds to
the dot starting with an electron in the ground state. Then this electron
tunnels to the right and an electron from the left tunnels to the excited
state, and then the electron tunnels from the excited state to the right
lead, and another electron tunnels from the same lead to the ground state,
ending at the same initial state but one electron transferred from left to
right. This probability amplitude interferes with  the amplitude of one
electron tunneling directly through the other arm from left to right. The
second  process (Fig.~\ref{fig:third_order}b) starts with the same initial
state, and involves an electron tunneling through the other arm to the right
lead, and then an electron from the right lead tunneling to the excited state,
while the ground state electron tunnels to the right lead. This amplitude,
which again involves one electron moving from left to right, interferes with
the amplitude where the dot electron tunnels to the right and an electron
from the left tunnels to the excited state. These two processes, which have
the same weight as they start from the same initial configuration,  involve
the exact same matrix elements, but effectively correspond to electron
traversing the AB ring in opposite directions, thus leading to the cancellation
of the term odd in magnetic field. Similar cancellation occurs for the
processes
starting with the QD in its exited state, Fig.~\ref{fig:third_order}c,d.

However, let us examine the process shown in Fig. \ref{fig:third_order}e.
The process that should cancel its asymmetric contribution is depicted in
Fig. \ref{fig:third_order}f. This latter process, however, does not
contribute to the current, as it describes electron backscattered into
the same lead. Thus, the contribution of the elastic process in
Fig. \ref{fig:third_order}e gives rise to AB oscillations odd in magnetic
field. Figs. \ref{fig:third_order}g,h provide an example of a similar
non-canceling inelastic process.

The distinctive feature of the processes in Figs. \ref{fig:third_order}e
and \ref{fig:third_order}g is that prior to electron transfer from left
to right, an electron is being excited to a state within the same lead.
When this part of the process is singled out as a one-particle amplitude
in the other process made up of the same elements and beginning from the
same initial state, Figs. \ref{fig:third_order}f and \ref{fig:third_order}h,
we obtain processes which only involve excitation within the same lead, and
thus do not contribute to the current, i.e. do not contribute to the measured
AB oscillations.

Since such a preliminary excitation is possible only when the QD is initially
in its excited state, whose population differs from zero only when $eV>\Delta$,
{\em breaking of the phase symmetry may happen only after the onset of
inelastic cotunneling}.

The asymmetric contribution to AB oscillations is of higher order in the
lead-to-lead coupling than the symmetric contribution. Thus, the asymmetry
should be weak everywhere, except the bias values where second order processes
vanish due to canceling contributions from different levels, i.e when phase
switching occurs \cite{PullerMeir}.  Overall, this means that \emph{the phase
of AB oscillations is not a monotonous function of bias: it is usually very
close to $0,\pi$, but deviates significantly from these values when phase
switching occurs.}

\paragraph{Discussion and comparison to the experiment}
Here we report calculations with a three level dot, similar to that used in
Ref.~\cite{PullerMeir}
in connection to the experiments of Ref.~\cite{Ensslin}: the levels have
alternating parity and different strength of coupling to the leads.

\begin{figure}[tbp]
  \includegraphics[width=3in]{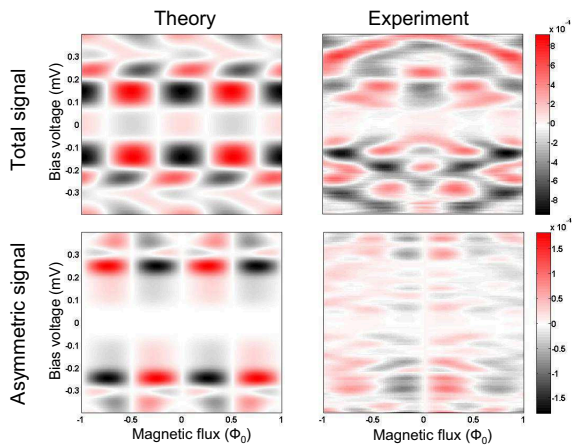}
  \caption{(color online) Colorplots of the differential conductance obtained
  from the theoretical model presented here (left panels), and from the
  experimental data of Ref. \onlinecite{Ensslin} (right panels). The upper
  and lower panels show respectively full and asymmetric components of the
  conductance.}\label{fig:exp_theor}

\end{figure}

The AB component of differential conductance obtained within the perturbation
framework
described above is shown in the upper left panel of Fig.~\ref{fig:exp_theor}.
One can see that the phase of the AB oscillations changes between $0$ and
$\pi$. In order to judge whether oscillations are strictly symmetric or not
we provide in the lower left panel of Fig.~\ref{fig:exp_theor} the colorplot
for the asymmetric component of AB oscillations extracted from the data shown
in the upper left. The right part of Fig. \ref{fig:exp_theor} presents
respectively total (upper panel) and asymmetric (lower panel) contributions
to AB oscillations as obtained from the experimental data of
Ref.~\cite{Ensslin}.

In both theoretical and experimental colorplots one can observe several
important features: (i) the phase of AB oscillations switches sharply
between values close to $0$ and $\pi$ \cite{Ensslin,PullerMeir};
(ii) in the figures showing total AB signal any significant asymmetry is
seen only in the regions corresponding to phase switching, e.g., close to
$V=\pm .2mV$  in the upper part of Fig.~\ref{fig:exp_theor};
(iii) the asymmetric component of AB oscillations is zero for bias below
the onset of inelastic cotunneling, but non-zero essentially everywhere
above this onset.

\begin{figure}[tbp]
  \includegraphics[width=3in]{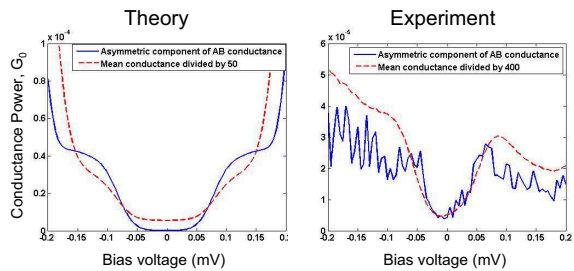}
  \caption{(color online) Power of asymmetric AB oscillations and differential
  conductance (rescaled) for theoretical model (left) and for experimental
  data (right). }\label{fig:power}

\end{figure}

In order to illustrate the last point we show in Fig.~\ref{fig:power}
the mean differential conductance through the interferometer together
with the power of the asymmetric component, calculated as
$P(V_{sd}) = \sqrt{ \int_{B_{min}}^{B_{max}}
dB G_{asym}^2 (B,V_{sd}) }/(B_{max}-B_{min})$, where $G_{asym} (B,V_{SD})$
is the asymmetric component of the differential conductance as a function
of magnetic field $B$ and bias voltage, $V_{sd}$. For the theoretical model
limits $B_{min}$ and $B_{max}$ are restricted to one period of AB
oscillations. At the onset of inelastic cotunneling the differential
conductance exhibits a jump, which is due to increase of the available
conductance processes. We see that the power of the asymmetric component
mimics the onset of inelastic cotunneling, which confirms our theoretical
predictions. The non-zero value of the asymmetric AB oscillations before
the onset of inelastic cotunneling in experimental data most likely results
from finite extension of the electron density throughout the device
(i.e. not all localized to QD). In this case the electric potential within
the device becomes a function of magnetic field, which leads to asymmetry
of AB oscillations \cite{Buttiker2}, which however grows smoothly with the
bias voltage \cite{electricAB}.

Although the asymmetric component makes the AB phase change continuously
between $0$ and $\pi$, most of the time the phase is very close to one of
these values. This is very different from smooth (almost linear) change of
the AB phase observed in AB rings without discrete levels structure
\cite{electricAB}. In this sense the most striking feature is the appearance
of the asymmetric AB oscillations only after the onset of inelastic
cotunneling.

In our theoretical figure the conductance odd in magnetic field is also odd
in bias (see also Ref.\onlinecite{Buttiker2}), a symmetry which is not
obeyed by the experimental results. Presence of a conductance component
odd in magnetic field, but even in bias indicates strong asymmetry of coupling
between the QD and the two leads. The lowest order terms in the expansion in
coupling strength are not sufficient in order to describe quantitatively the
experimental asymmetry. Therefore we limited our theoretical calculation to
the symmetric structure, where we chose the parameters that make theoretical
and experimental result match for positive bias.


\paragraph{Conclusion.} We addressed breaking of phase symmetry in a quantum
dot AB interferometer in cotunneling regime. We showed that AB oscillations
remain strictly symmetric up to the onset of inelastic cotunneling, and
discussed the processes responsible for breaking of the phase symmetry
above this onset. As asymmetric component of AB oscillations is of higher
order in lead-to-lead tunneling than the symmetric one, the AB phase
remains close to values $0$ and $\pi$.  The exception are the bias values
where phase switching occurs, and the asymmetric component of AB oscillations
becomes dominant. Altogether this results in AB phase changing sharply but
continuously between values $0$ and $\pi$.  We show that our theoretical
findings are in excellent agreement with the experimental data of
Ref.~\onlinecite{Ensslin}.

We thank Y. Gefen, V. Kashcheyevs, T. Aono and M. Khodas for useful
discussions.
We are grateful to O. Entin-Wohlman and A. Golub for valuable comments.
This work was supported in part by the ISF and BSF. V.P. is partially
supported by Pratt Fellowship.


\begin{thebibliography}{99}
\bibitem{onedot} In the experiment each arm contained a dot. However,
the dot with the larger level spacing does not contribute to the multiple
level transport and thus may be considered as a tunneling barrier and is
replaced in the model by direct tunneling between the leads.
\bibitem{Supplement} We assume single-channel leads and include the
magnetic field explicitely only via Aharonov-Bohm phase. A more general
analysis can be performed based on the properties of the tunneling matrix
elements in respect to reversal of magnetic field.
\end{thebibliography}
\end{document}